\pdfoutput=1 
\documentclass[12pt]{article}

\usepackage{graphicx}
\usepackage{amsmath}
\usepackage{amssymb}
\usepackage{lineno}
\usepackage{url}
\usepackage{xspace,multicol}
\usepackage{siunitx}
\usepackage{subcaption}
\usepackage{color, colortbl}
\usepackage{units}
\usepackage{ragged2e}
\usepackage{array}
\usepackage{tabularx}
\usepackage{authblk}
\usepackage{heppennames}
\usepackage{feynmp}
\usepackage{libertine}
\usepackage{textgreek}
\newcommand{\geant}{\textsc{Geant4}\xspace}
\newcommand{\murm}{%
  \ifmmode
    \mathchoice
        {\hbox{\normalsize\textmu}}
        {\hbox{\normalsize\textmu}}
        {\hbox{\scriptsize\textmu}}
        {\hbox{\tiny\textmu}}%
  \else
    \textmu
  \fi
}

\usepackage{authblk}

\begin{document}


\title{A Study of the Impact of Muons from the Beam Delivery System on the\\SiD Performance\vspace*{0.3cm}\\{\normalsize Talk presented at the\\International Workshop on Future Linear Colliders (LCWS2016)\\Morioka, Japan, 5-9 December 2016. C16-12-05.4.}}

\author[1,2]{Anne Sch\"utz}
\author[3]{Lewis Keller}
\author[3]{Glen White}

\affil[1]{\normalsize Karlsruhe Institute of Technology (KIT), Department of Physics, Institute of Experimental Nuclear Physics (IEKP), Wolfgang-Gaede-Str. 1, 76131 Karlsruhe}
\affil[2]{\normalsize Deutsches Elektronen-Synchrotron (DESY), Notkestr. 85, 22607 Hamburg}
\affil[3]{\normalsize SLAC National Accelerator Laboratory, 2575 Sand Hill Rd, Menlo Park, CA 94025}

\maketitle


\begin{abstract}
To suppress the muon background arising from the Beam Delivery System (BDS) of the International Linear Collider (ILC), and to hinder it from reaching the interaction region, two different shielding scenarios are under discussion: five cylindrical muon spoilers with or without an additional magnetized shielding wall.
Due to cost and safety issues, the scenario preferred by the Machine-Detector-Interface (MDI) group is to omit the shielding wall, although omitting it also has disadvantages.
To support the decision making for the muon shielding, the impact of the muons from the two different shielding scenarios was studied in a full Geant4 detector simulation of the SiD detector, one of two proposed detectors for the ILC. 
Input to this study is the muon background created by the beam traveling through the BDS, which was simulated with MUCARLO.\cite{Mucarlo, MuonBkg_05TeV, MuonBkg_1TeV}
\end{abstract}

\section{Introduction}
\label{sec:introduction}

\begin{figure}
    \centering
    \begin{subfigure}[b]{0.53\textwidth}
        \includegraphics[width=\textwidth]{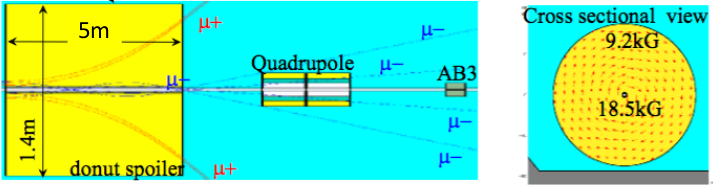}
        \caption{The magnetized iron spoilers}
	\label{fig:spoilers}
    \end{subfigure}
    \begin{subfigure}[b]{0.42\textwidth}
        \includegraphics[width=\textwidth]{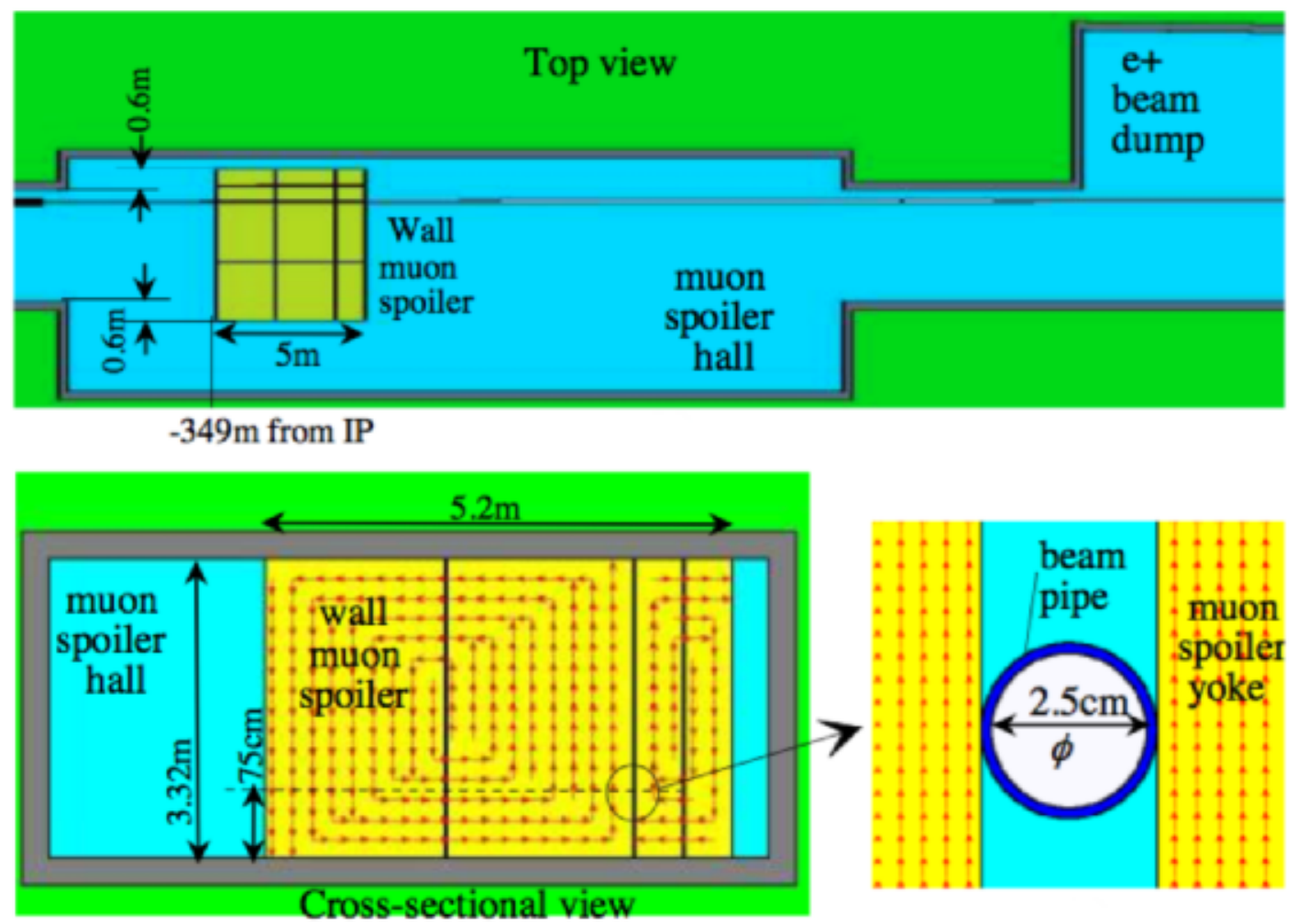}
        \caption{The magnetized iron wall}
        \label{fig:wall}
    \end{subfigure}
    \caption[Schematic drawings of the shielding systems]{
    Schematic drawings of the magnetized cylindrical spoilers and the magnetized wall.
    The spoilers have a radius of \unit{70}\,{cm} and a length of \unit{5}\,{m}, while the wall is about \unit{5}\,{m} in length and width and almost fills the entire tunnel in height.
    }
    \label{fig:Spoilers_Wall}
\end{figure}

\begin{figure}
    \centering
    \includegraphics[width=0.7\textwidth]{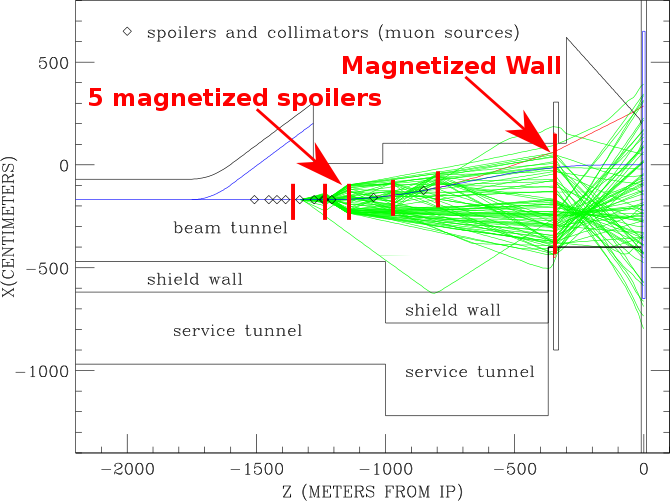}
    \caption[BDS tunnel with the spoiler and wall positions]{
    xz-view of the electron line of the BDS with the magnetized spoilers and the magnetized wall being depicted at their particular locations.
    The green and red lines represent the tracks of the positively and negatively charged muons respectively, originating at this specific point along the beam line.
    The reason that almost entirely positive muons are drawn is that only tracks that reach the detector are displayed.
    As the spoiler polarities are set to defocus muons with the same charge as the beam charge, the \textmu\textsuperscript{-} are defocused and therefore absorbed in the tunnel walls before hitting the detector.
    }
    \label{fig:Spoilers_Wall_Locations}
\end{figure}

The muon background from the Beam Delivery System (BDS) arises from the the beam halo hitting the material along the beam line, e.g. the beam collimator systems.
Therefore, the muons are created along the BDS and travel towards the interaction region.
To prevent the muons from reaching the detectors, a study was performed to decide which shielding system would be effective and reasonable to be integrated in the BDS.
Several different shielding scenarios have already been under discussion but did not fulfill the requirements.
The two cases currently discussed and presented in this note are both yielding muon rates below 5 muons per bunch crossing at the interaction region.
In the first scenario five cylindrical magnetized spoilers are installed at different positions along the beam line.
It will be referred to as the ``5 Spoilers'' scenrario.
The second scenario adds a magnetized iron wall close to the interaction region, and is therfore called the ``5 Spoilers + Wall'' scenrario.
Both, the spoilers and the wall are shown in schematic drawings in Figure~\ref{fig:Spoilers_Wall}.
Their location in the BDS can be seen in Figure~\ref{fig:Spoilers_Wall_Locations}.
The five spoilers are located at the following positions along the BDS tunnel: \unit{802.5}\,{m}, \unit{975.5}\,{m}, \unit{1145.5}\,{m}, \unit{1234.5}\,{m}, and \unit{1358.5}\,{m}.
For the second shielding scenario, the wall would be additionally installed at \unit{400}\,{m} away from the interaction point.\cite{Lewis}

Due to the size of the wall, which has safety issues and implies additional costs, the MDI group prefers the first scenario without the wall.
What has to be kept in mind on the other hand is that the wall reduces not only the muon rate but also shields the neutron and photon background created by the machine.
Additionally, the wall acts as a tertiary containment device which prevents potentially harmful levels of muon fluxes in the event that the primary Personnel Protection Systems fail and the main beam hits a beam stopper in the BDS tunnel.
Having this shielding wall therefore means the access to the interaction region, i.e. the access to the detector in the garage position, would be permitted when the beam is on.
\\
To facilitate the decision of whether the wall is needed or not, the detector groups were asked to study the impact of the muons on the performance of the detectors.
This note shows the results of the study done for the SiD detector\cite{SiDLOI,TDR4}.\\
In the current SiD detector design, the used sensors have a buffer depth of four, i.e. the sensors can store four hits in their buffers.
Since these buffers are read out after every ILC bunch train only, which corresponds to 1312 bunch crossings, the detector occupancy\footnote{The occupancy describes the fraction of detector cells that have been hit a certain amount of times.} from backgrounds has to be kept as low as possible.
With a tracker occupancy of 10\textsuperscript{-3} being considered acceptable, the SiD occupancy caused by the muon background has to be seen in the context of all occurring backgrounds.
This will have an impact on which muon shielding system the SiD group prefers.

\section{The Simulation of the Muon Background with MUCARLO}
\label{MUCARLO}

The simulation code MUCARLO\cite{MuonBkg_05TeV, MuonBkg_1TeV} is based on Fortran, and was originally written by Gary Feldman.
Over the years it has been expanded, and is used in several studies, from the study of muon shielding designs for radiation protection, to fixed target experiments at SLAC and muon background simulation studies for the Next Linear Collider (NLC) and the ILC\cite{MuonBkg_05TeV, MuonBkg_1TeV}.\\
For the presented study, the Technical Design Report (TDR) baseline machine parameters for the ILC-500GeV are used for simulating the beam interacting with the BDS geometry and the muon collimation system.
The muons are produced in interactions of the beam halo with material in the beam lines, in which the predominant interaction is the Bethe-Heilter process:
\textgamma + Z $\rightarrow$ Z' + \murm\textsuperscript{+}\murm\textsuperscript{-}\\
The muon production by direct annihilation of the positrons with atomic electrons is also taken into account.\cite[sec. 2]{Mucarlo}
For tracking the beam halo, the tool TURTLE\cite{Turtle} is used.\\
The results from the MUCARLO simulations can be seen in Table~\ref{tab:MuonRates}, listing the number of muons reaching the interaction region for the two shielding scenarios and for the case of not having any muon shielding system.
The calculated muon rate is based on a halo population of 10\textsuperscript{-3}, which is more than ten times larger than expected from ring scattering calculations.
This estimation corresponds to the worst halo measured at the Stanford Linear Collider (SLC), and is therefore used as a worst-case scenario.

\begin{table}
\caption{MUCARLO results: The number of muons hitting a detector with radius of \unit{6.5}{m} in the different shielding scenarios.}
\label{tab:MuonRates}
\centering
\begin{tabularx}{\textwidth}{ll}
\hline\hline
\textbf{Scenario} & \textbf{Number of muons in a detector with 6.5m radius}\\
\hline
\cline{1-2}
\hline
 No Spoilers & 130 muons/bunch crossing\\
 5 Spoilers& 4.3 muons/bunch crossing\\
 5 Spoilers + Wall & 0.68 muons/bunch crossing\\
\hline\hline
\end{tabularx}
\end{table}
\section{The simulation of muons in the SiD detector}
\label{Detector}

\begin{table}
\caption[Time structure of the ILC-500GeV beam.]{Time structure of the ILC-500GeV beam. The bunch spacing is the time between two bunches, whereas the collision rate defines with which frequency the bunch trains are colliding.}
\label{tab:Beam_TimeStructure}
\begin{tabularx}{\textwidth}{lll}
\hline\hline
\textbf{Number of bunches per train} & \textbf{Bunch spacing} &  \textbf{Collision rate} \tabularnewline
\hline
1312 & \unit{554}\,{ns} & \unit{5}\,{Hz}\tabularnewline
\hline\hline
\end{tabularx}
\end{table}

\begin{figure}
    \centering
    \includegraphics[height=0.4\textheight]{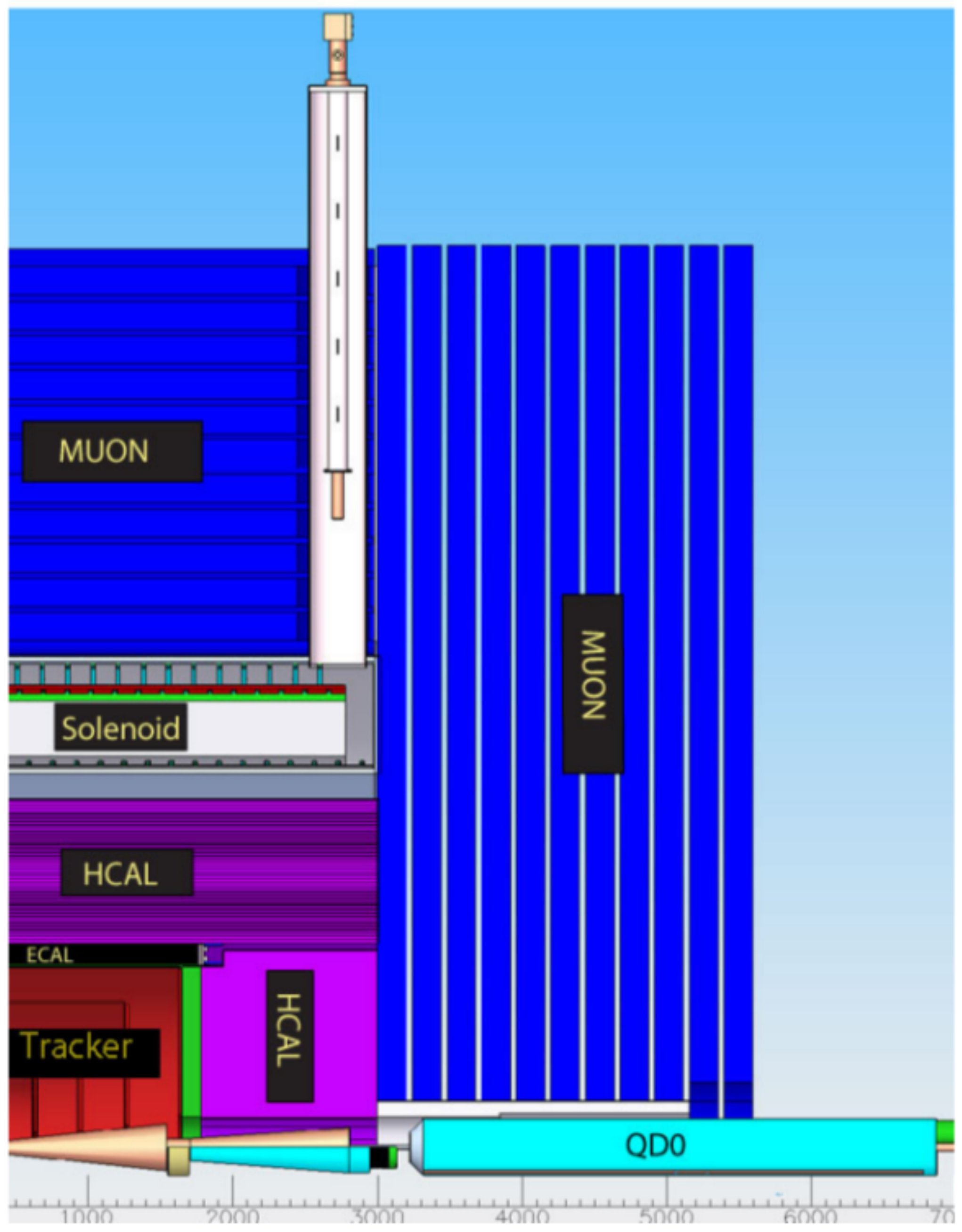}
    \caption[SiD detector cross sections]{
    Cross sections of the SiD detector in the xy plane (left), and the xz plane (right).
    The schematics show the SiD concept design from the Techical Design Report~\cite[p. 59]{TDR4}.
    }
    \label{fig:SiD}
\end{figure}

\begin{figure}
    \centering
    \begin{subfigure}[b]{0.49\textwidth}
    \begin{center}
        \includegraphics[height=0.3\textheight]{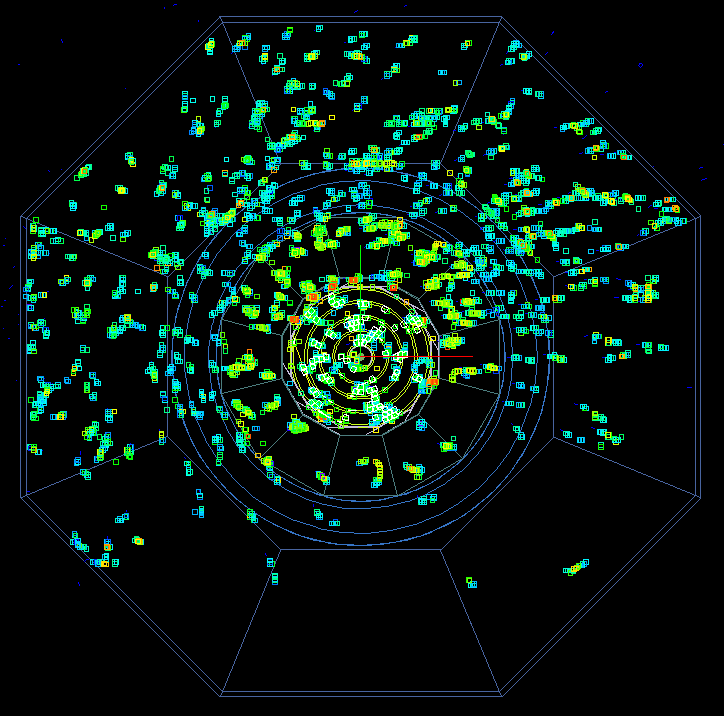}
        \caption{xy-view, ``5 Spoilers + Wall''}
	\label{fig:xy_5SpoilersWall}
    \end{center}
    \end{subfigure}
    \begin{subfigure}[b]{0.49\textwidth}
    \begin{center}
        \includegraphics[height=0.3\textheight]{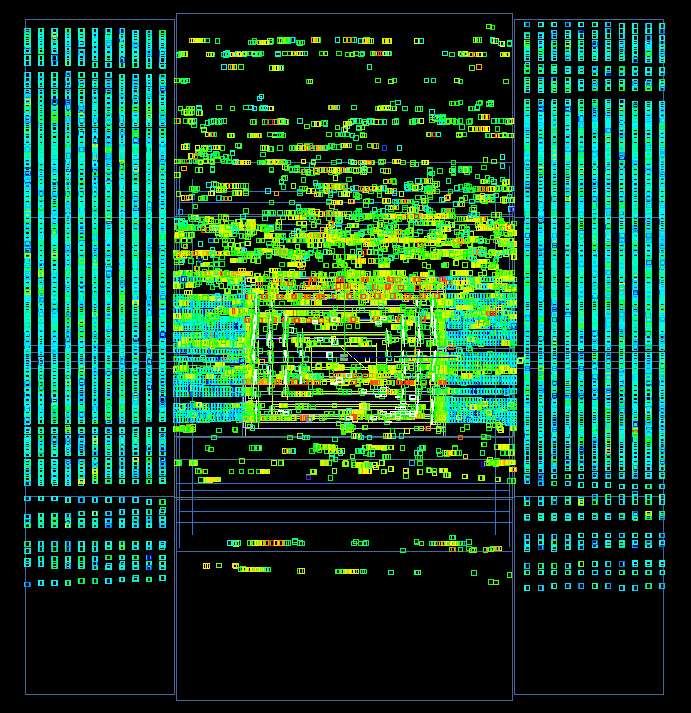}
        \caption{zy-view, ``5 Spoilers + Wall''}
        \label{fig:zy_5SpoilersWall}
    \end{center}
    \end{subfigure}\\
    \begin{subfigure}[b]{0.49\textwidth}
    \begin{center}
        \includegraphics[height=0.3\textheight]{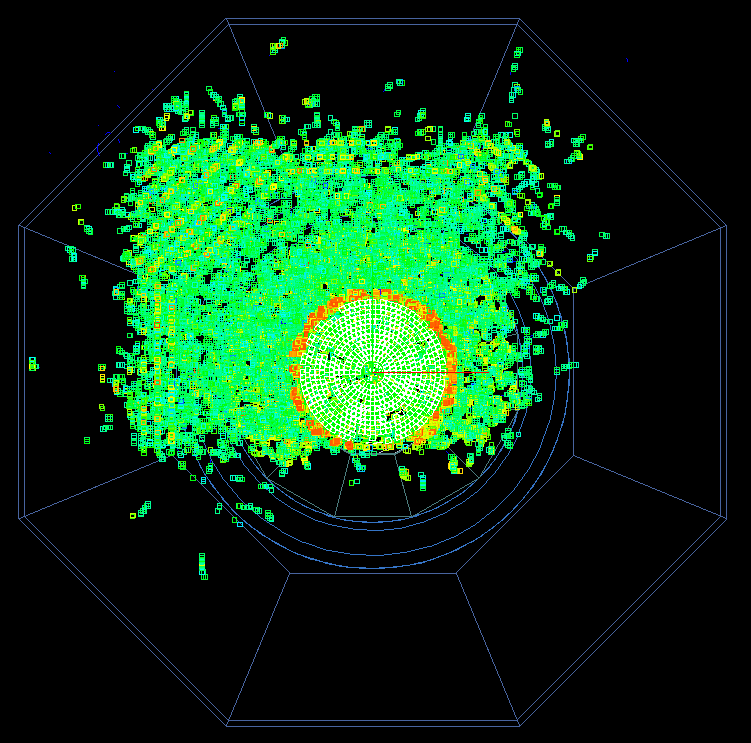}
        \caption{xy-view, ``5 Spoilers''}
	\label{fig:xy_5Spoilers}
    \end{center}
    \end{subfigure}
    \begin{subfigure}[b]{0.49\textwidth}
    \begin{center}
        \includegraphics[height=0.3\textheight]{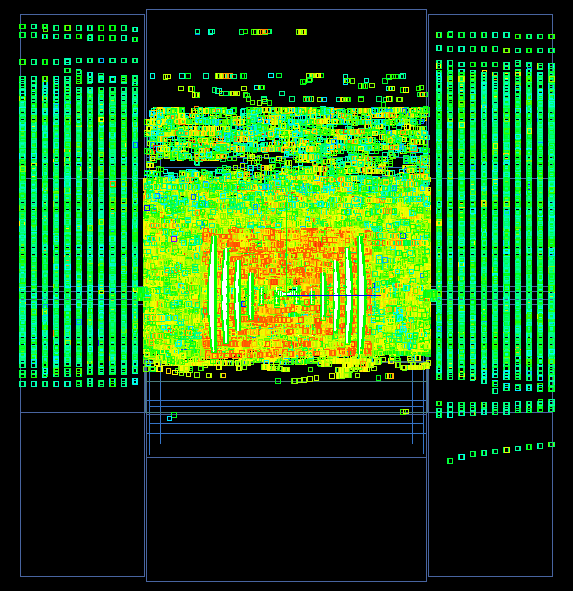}
        \caption{zy-view, ``5 Spoilers''}
        \label{fig:zy_5Spoilers}
    \end{center}
    \end{subfigure}
    \caption[Event displays of muons in SiD]{
    Event displays in the xy- and zy-view of the muons from the ``5 Spoilers + Wall'' and ``5 Spoilers'' scenario in the SiD detector.
    Figures a) and b) show hits from 515 muons with the magnetized wall, and c) and d) show hits from 2961 muons without the wall.
    The number of muons corresponds to the number of muons accumulated over one ILC bunch train (1312 bunch crossings), but in both cases the muons come from the positron line of the BDS only.
    Hence, one has to imagine roughly double the amount of hits in the SiD detector to get the full picture of all hits from muons from one bunch train.
    }
    \label{fig:WIRED4}
\end{figure}

\begin{figure}
    \centering
    \includegraphics[height=0.3\textheight]{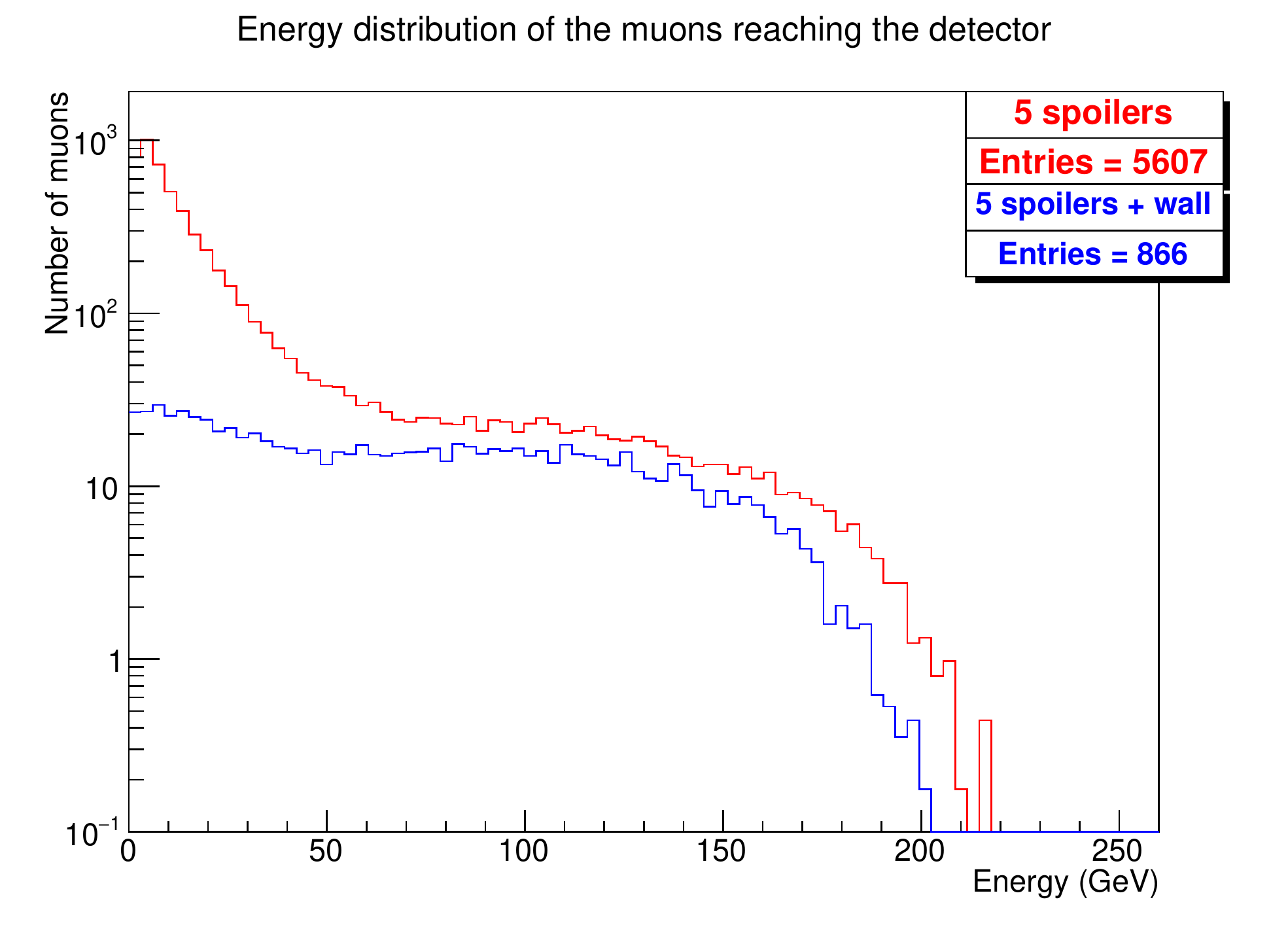}
    \caption[Energy distribution of muons from the two shielding scenarios]{
    The energy distributions of the muons from one ILC bunch train (1312 bunch crossings) for both scenarios, ``5 Spoilers'' (red) and ``5 Spoilers + Wall'' (blue), show that magnetized wall deflects and stops the low energy muons.
    The peak for low energies is therefore missing in the second scenario, and the whole distribution is shifted towards lower energies.
    }
    \label{fig:muon_energy}
\end{figure}

\begin{figure}
    \centering
    \includegraphics[height=0.32\textheight]{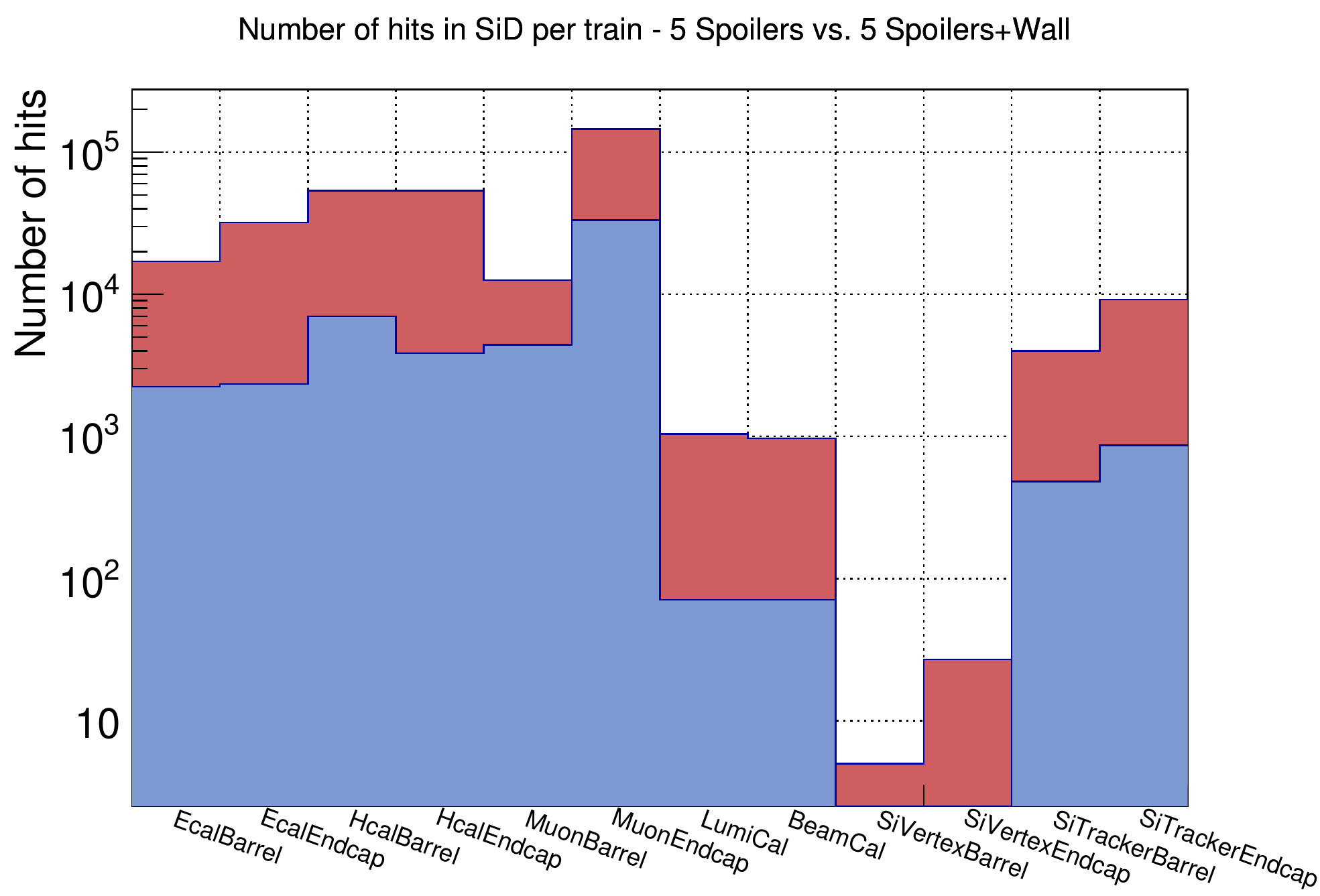}
    \caption[Hit number distribution in the SiD subdetectors]{
    The hit number distributions of the muon hits in the SiD subdetectors from one ILC bunch train (1312 bunch crossings).
    The two scenarios are colored in red (``5 Spoilers'') and blue (``5 Spoilers + Wall'') as before.
    The number of hits is proportional to the effective detector area.
    Since the muons travel horizontally through the detector from one side to the other, the detector with the biggest effective area perpendicular to the muon incidence has the highest number of hits.
    }
    \label{fig:hit_distribution}
\end{figure}

In the first step of simulating the muons in the SiD detector, event displays were made with WIRED4\cite{Wired4}.
The xy- and zy-views of these event displays can be seen for both shielding scenarios in Figure~\ref{fig:WIRED4}.
The event displays show the hits in the SiD detector for one ILC bunch train, which consists of 1312 bunches.
The time structure of the ILC bunch trains is shown in Table~\ref{tab:Beam_TimeStructure}.
First of all, without the magnetized wall as a last shielding mechanism about six times more muons enter the interaction region.
In these plots, as in all following, the shown distributions are from muons from one ILC bunch train, since the SiD sensor buffers will be read out after every train only.\\
Second, the muons travel through the detector horizontally which could be taken advantage of for the alignment of the subdetectors.
The spatial distributions of the muons in both scenarios are shifted towards the top and the left.
The reason for the top-bottom and left-right asymmetry is the shielding ability of the tunnel walls and floor.
Since the detector hall is set into the floor, and the BDS system is curved, only the muons being emitted in such a way that they get through the tunnel opening to the detector hall hit the detector.
Finally, the most significant difference between the two scenarios is the broad spatial distribution for the scenario with the wall in comparison to the distinct distribution in the ``5 Spoilers'' scenario.
The magnetized wall deflects the muons and also stops the low energy ones, so that the muon rate is reduced and the muons are additionally distributed over the whole detector area.
Figure~\ref{fig:muon_energy} shows the muon energy distributions for both shielding scenarios.
The difference in the spatial distributions explains the different numbers of hits in the single subdetectors for the two scenarios, as shown in Figure~\ref{fig:hit_distribution}.
With the muon endcaps having the largest effective detector area, the number of hits is the highest in this subdetector.
Table~\ref{tab:KeyParametersSiD} lists the different components of the SiD detector with their physical sizes, the technology used, and the sizes of the readout cells.


\begin{table}
\caption{Key parameters of the baseline SiD design, including the measurements of the subdetectors, their readout cell dimensions and the buffer depth. The given readout cell dimension are the pixelation cell sizes used for the full detector \geant simulation.}
\label{tab:KeyParametersSiD}
\begin{tabular}{>{\raggedright}p{1.5cm}>{\raggedright}p{2.5cm}>{\raggedright}p{2.4cm}>{\centering}p{1cm}>{\raggedright}p{1.2cm}>{\raggedright}p{1.2cm}>{\raggedright}p{1.2cm}}
\hline\hline
\textbf{SiD Barrel} & \textbf{Technology} & \textbf{Readout cell dimensions [mm$^2$]} & \textbf{Buffer depth} & \textbf{Inner radius [cm]} & \textbf{Outer radius [cm]} & \textbf{z extent [cm]} \tabularnewline
\hline
Vertex detector & Silicon pixels & 0.05 x 0.05 & 4 & 1.4 & 6.0 & $\pm 6.25$ \tabularnewline
Tracker & Silicon strips & 0.05 x 0.05 & 4 & 21.7 & 122.1 & $\pm 152.2$ \tabularnewline
ECAL & Silicon pixels-W & 3.5 x 3.5 & 4 & 126.5 & 140.9 & $\pm 176.5$ \tabularnewline
HCAL & RPC-steel & 10 x 10 & 4 & 141.7 & 249.3 & $\pm 301.8$ \tabularnewline
Solenoid & 5 T SC & - & - & 259.1 & 339.2 & $\pm 298.3$ \tabularnewline
Flux return & Scintillator-steel & 30 x 30 & 4 & 340.2 & 604.2 & $\pm 303.3$ \tabularnewline
\hline
\textbf{SiD Endcap} & \textbf{Technology} & \textbf{Readout cell dimensions [mm$^2$]} & \textbf{Buffer depth} & \textbf{Inner z [cm]} & \textbf{Outer z [cm]} & \textbf{Outer radius [cm]} \tabularnewline
\hline
Vertex detector & Silicon pixels & 0.05 x 0.05 & 4 & 7.3 & 83.4 & 16.6 \tabularnewline
Tracker & Silicon strips & 0.05 x 0.05 & 4 & 77.0 & 164.3 & 125.5 \tabularnewline
ECAL & Silicon pixel-W & 3.5 x 3.5 & 4 & 165.7 & 180.0 & 125.0 \tabularnewline
HCAL & RPC-steel & 10 x 10 & 4 & 180.5 & 302.8 & 140.2 \tabularnewline
Flux return & Scintillator/steel & 30 x 30 & 4 & 303.3 & 567.3 & 604.2 \tabularnewline
LumiCal & Silicon-W & 3.5 x 3.5 & 4 & 155.7 & 169.55 &  20.0 \tabularnewline
BeamCal & Semicond.-W & 3.5 x 3.5 & 4 & 326.5 & 344 & 14.0 \tabularnewline
\hline\hline
\end{tabular}
\end{table}

\begin{figure}
    \centering
    \begin{subfigure}[b]{0.49\textwidth}
    \centering
        \includegraphics[height=0.265\textheight]{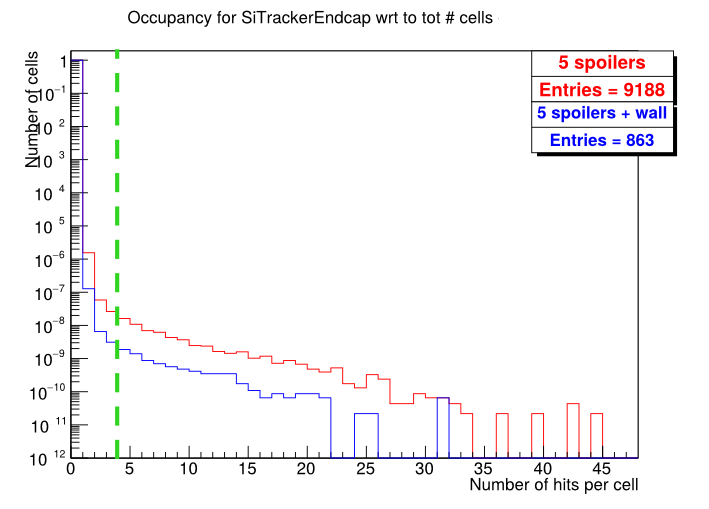}
        \caption{Tracker endcap occupancy}
	\label{fig:SiTracker_Occupancy}
    \end{subfigure}
    \begin{subfigure}[b]{0.49\textwidth}
    \centering
        \includegraphics[height=0.265\textheight]{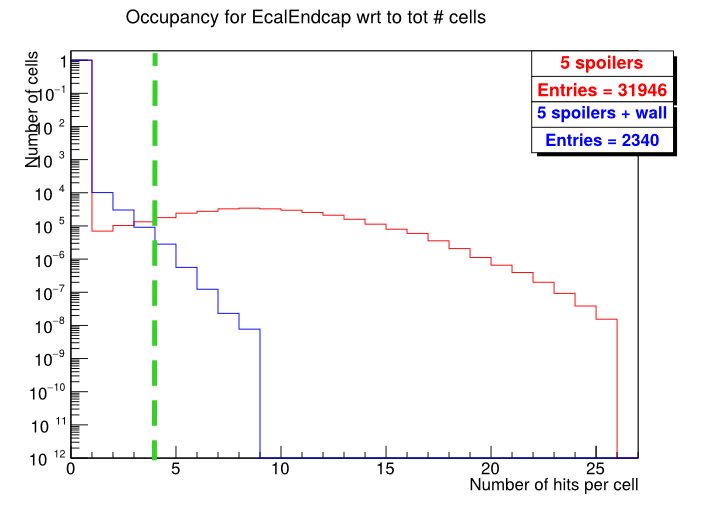}
        \caption{ECAL endcap occupancy}
        \label{fig:Ecal_Occupancy}
    \end{subfigure}\\
    \begin{subfigure}[b]{0.49\textwidth}
    \centering
        \includegraphics[height=0.265\textheight]{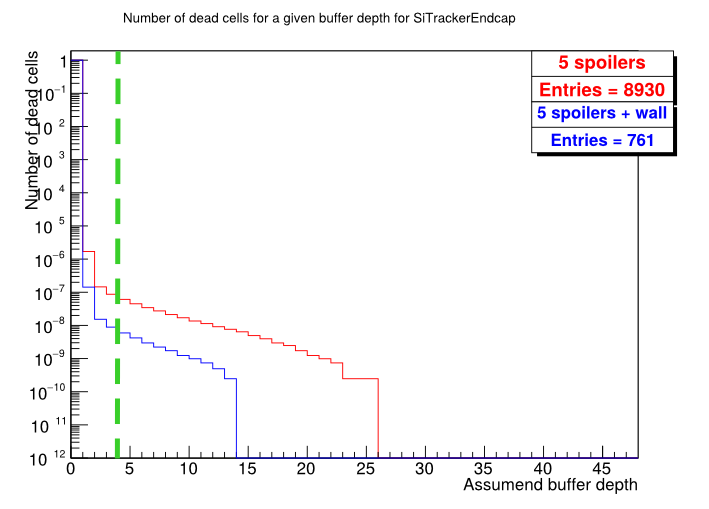}
        \caption{Dead cells in the tracker endcaps}
	\label{fig:SiTracker_DeadCells}
    \end{subfigure}
    \begin{subfigure}[b]{0.49\textwidth}
    \centering
        \includegraphics[height=0.265\textheight]{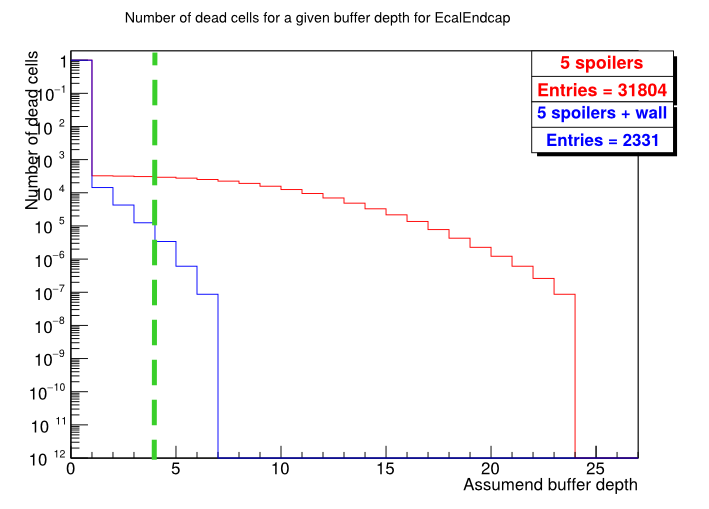}
        \caption{Dead cells in the ECAL endcaps}
        \label{fig:Ecal_DeadCells}
    \end{subfigure}
    \caption[Muon occupancy in the tracker endcaps and ECAL endcaps]{
    Figures ~\ref{fig:SiTracker_Occupancy} and \ref{fig:Ecal_Occupancy} show for both shielding scenarios the muon occupancy in the tracker endcaps and the ECAL endcaps, i.e. the fraction of all cells that are hit a certain number of times.
    The plots are normalized by the total number of cells in this subdetector.\\
    Figures~\ref{fig:SiTracker_DeadCells} and \ref{fig:Ecal_DeadCells} show in comparison the number of dead cells which is the result of the occupancy and the buffer depth of the sensors.
    For a given buffer depth, all cells with hit numbers greater or equal than the buffer depth are ``dead'', and therefore blind to all following hits.
    Therefore, in the hypothetical case of a buffer depth of 0, all cells are dead.\\
    In all plots, the green dashed line represents the buffer depth of the current sensor design.
    }
    \label{fig:Occupancy_DeadCells}
\end{figure}

\newpage
Cells are declared ``dead'' when they are hit by more tracks than the buffer depth allows, as any hits beyond the buffer depth limit can not be recorded.
Figure~\ref{fig:Occupancy_DeadCells} shows the occupancy plots and the number of dead cells resulting from the occupancy for both scenarios, and for two different subdetectors: the tracker endcaps and the ECAL endcaps. 
The occupancy in the tracker endcaps (Figure~\ref{fig:SiTracker_Occupancy}) is very low for both scenarios.
Only 10\textsuperscript{-9} - 10\textsuperscript{-7} of all cells get four hits.
The resulting number of dead cells in the tracker endcaps (Figure~\ref{fig:SiTracker_DeadCells}) shows that for a buffer depth of four only 10\textsuperscript{-8} of all cells get four or more hits and therefore reach the buffer limit in the ``5 Spoilers + Wall''.
The ``5 Spoilers'' case would do only one order of magnitude better.\\
But the muon background yields a much higher occupancy in the ECAL endcaps which is shown in Figures~\ref{fig:Ecal_Occupancy} and \ref{fig:Ecal_DeadCells}.
The reason for the occupancy being higher than in the tracker endcaps for both cases is simply due to the bigger effective detector area as explained before.
Despite that and the fact that the ``5 Spoilers + Wall'' case is better by an order of magnitude (when looking at a buffer depth of four), the occupancy is still at a level of only about 10\textsuperscript{-6} - 10\textsuperscript{-5}.
The interesting fact is that the ``5 Spoilers'' case shows up to 27 hits per cell with a roughly constant occupancy for all buffer depths.
This leads to dead cell distributions which are vastly different.
For an assumed buffer depth of four, the total number of dead cells is different by about two orders of magnitude when comparing the two shielding scenarios.
In the ``5 Spoilers'' case, 10\textsuperscript{-4} - 10\textsuperscript{-3} cells would have reached the buffer limit regardless of which buffer depth was chosen for the sensor design.
The cause of this distribution is the spatial distribution of the muons in the ``5 Spoilers'' case: there are many more muons hitting the detector, and additionally all concentrated on a small area of the detector.

\begin{figure}
    \centering
    \includegraphics[width=0.7\textwidth]{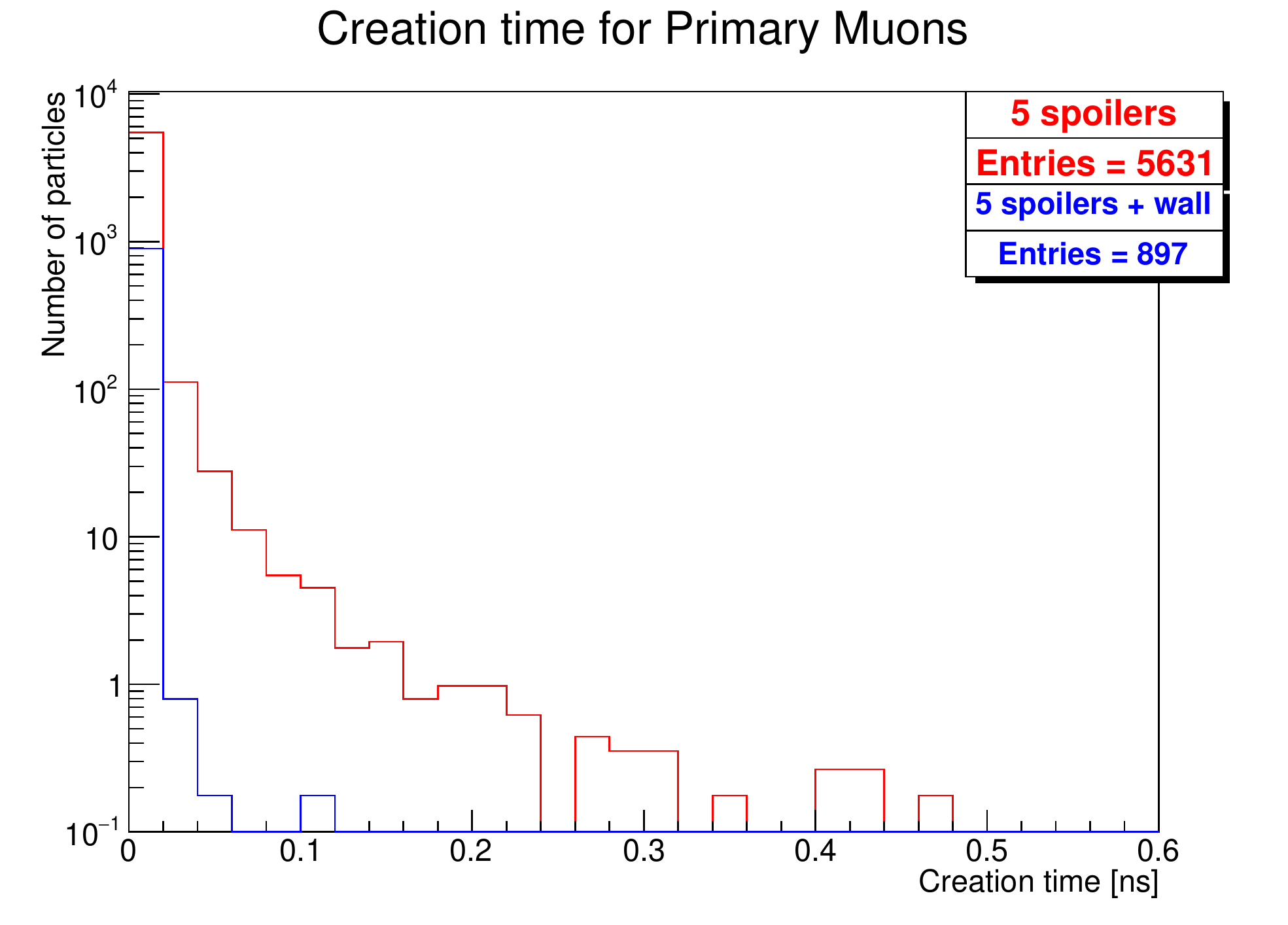}
    \caption[Creation time of the muons]{
    The creation time of the muons from the two shielding scenarios for a full bunch train.
    All of the muons from the BDS are created up to \unit{0.5}\,{ns} after the bunch passing the material, whereas the lower energy muons, which have a broader creation time, do not reach the detector in the ``5 Spoilers + Wall'' case.
    }
    \label{fig:Creation_time}
\end{figure}

Finally, also the timing of the muons with respect to the bunch crossing was studied.
All of the muons from the BDS are created up to \unit{0.5}\,{ns} after the bunch passing the material, as can be seen in Figure~\ref{fig:Creation_time}.
Although the muons are created instantaneously, it takes a long time for them to hit certain subdetectors, such as the inner lying ECAL.
The muons have to travel through the outer subdetectors before they reach the ECAL endcaps, which takes about \unit{20}\,{ns}.
After roughly another \unit{20}\,{ns} the second endcap has been reached, so that hits in the ECAL endcaps can be registered several tens of nanoseconds after the bunch crossing .
The muons also produce shower particles when passing through the whole detector material.
The low energy shower particles then hit the ECAL endcaps even later than the primary muons.

\begin{figure}
    \centering
    \begin{subfigure}[b]{0.49\textwidth}
    \centering
        \includegraphics[height=0.26\textheight]{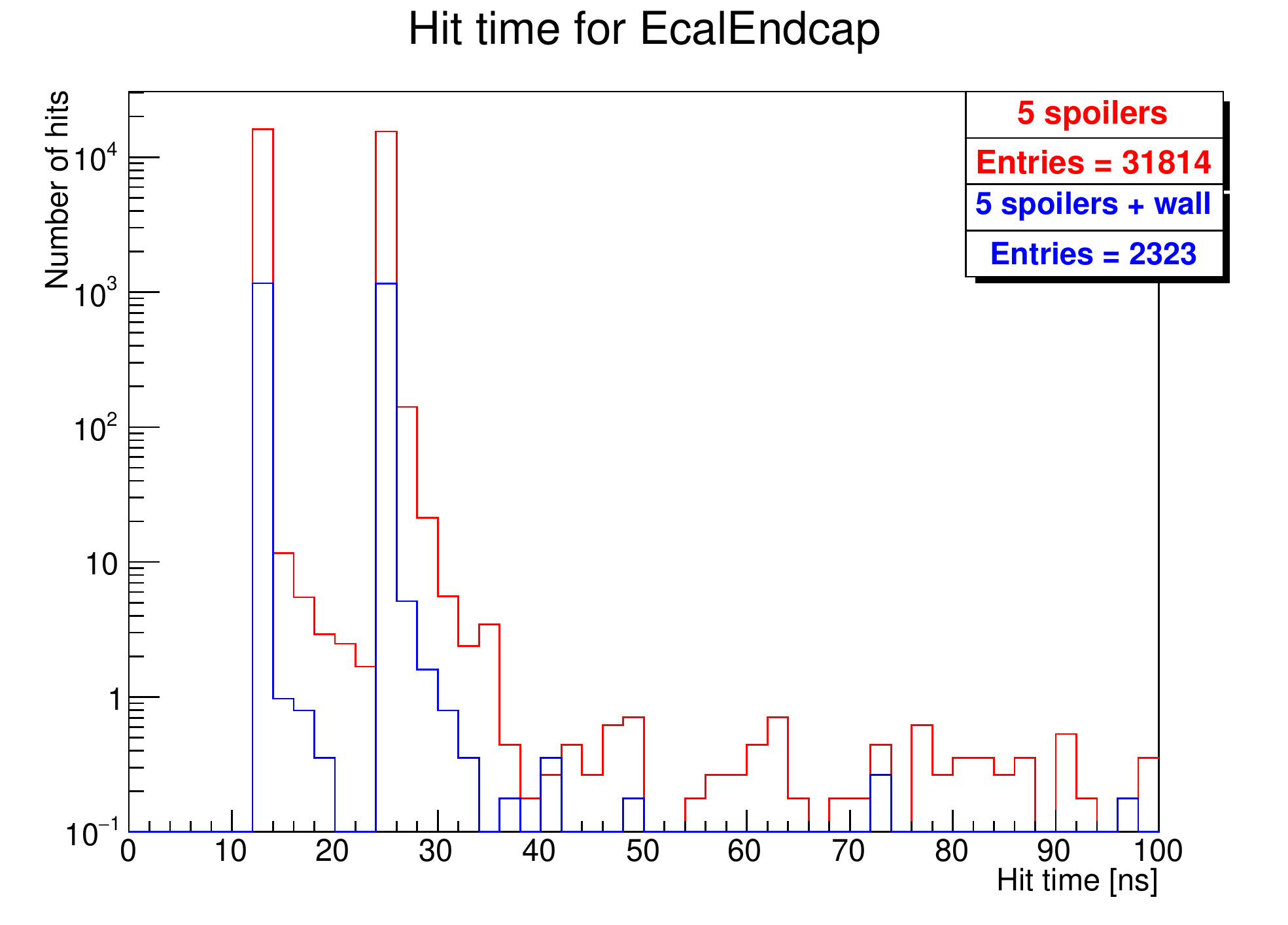}
        \caption{Hit time of the primary muons}
	\label{fig:hittime_ECAL}
    \end{subfigure}
    \begin{subfigure}[b]{0.49\textwidth}
    \centering
        \includegraphics[height=0.26\textheight]{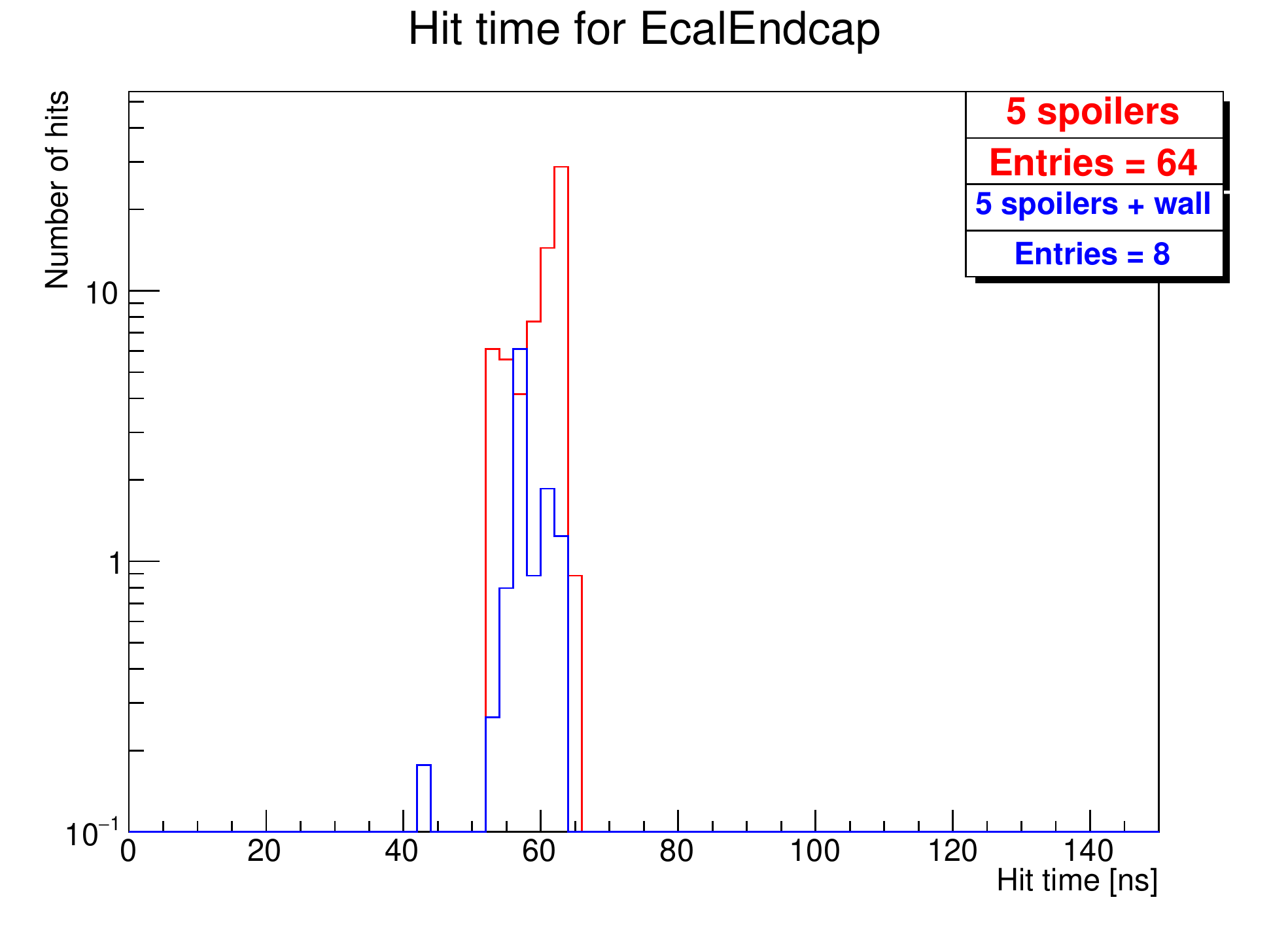}
        \caption{Hit time of the shower particles}
        \label{fig:hittime_shower_ECAL}
    \end{subfigure}
    \caption[Hit time distributions in the ECAL endcaps]{
    The distributions of the hit time of the primary muons and the shower particles hitting the ECAL endcaps.
    For both shielding scenarios, the time distributions are similar.
    The primary muons leave hits between about 23 and \unit{50}\,{ns} after the bunch crossing, since they have to travel through the whole detector before they reach the ECAL endcaps.
    The shower particles on the other hand hit the ECAL endcaps about \unit{60}\,{ns} after the crossing.
    }
    \label{fig:Hit_time}
\end{figure}
\section{Summary, Conclusions, and Outlook}

To facilitate the decision of whether a magnetized wall is necessary in addition to the five muon spoilers to guarantee acceptable muon shielding, the muons from the MUCARLO simulation have been simulated in a full SiD detector simulation.
In particular, the differences between the two shielding scenarios, and in both cases the impact of the muon background on the SiD occupancy have been studied.
The simulation showed that low energy muons from the Beam Delivery System are stopped or deflected by the magnetized wall, because of which the muon rate is reduced by a factor of about six.
Additionally, the spatial distributions of the muons in the two scenarios are quite different due to the magnetized wall, which deflects the muons and spreads them over the whole detector area.
Because of this, the number of hits in the different subdetectors are proportional to the effective detector areas.\\
Regarding the timing of the muons, it was shown that the primary muons are created instantaneously after the bunch halo passes through the material of the beam line.
Despite that the hit time distributions of the primary muons and the produced shower particles made clear that the detector registers hits up to about \unit{100}\,{ns} after the bunch crossing.\\
The most important objective of this study was the study of the SiD occupancy and the number of dead cells caused by the muon background.
Although the occupancy in the vertex and tracker detectors is way below the critical value, the occupancy in the ECAL endcaps for example almost reaches this critical limit for the shielding scenario without the magnetized wall.
It was found out that the reason for this is not only the higher muon rate in this scenario, but also the spatial distribution of the muons which are concentrated in a small area.\\
Overall, with the shown evaluation of the muons from the current MUCARLO simulations, the SiD group prefers to have the magnetized wall, in order to keep the occupancy from the backgrounds as low as possible.
It has to be kept in mind that the estimation of the muon rates is based on a worst-case scenario, in which the interacting beam halo population is more than ten times higher than normally expected.
This means that the muon occupancy in the SiD detector should be overestimated, and that the magnetized wall may then not be necessary.
On the other hand, a counter argument is the fact that, for an ILC upgrade to \unit{1}\,{TeV} center-of-mass energy, the considerably higher muon rates might make a muon wall necessary anyway.
As it is also a tertiary containment device against not only muons, but also photons and neutrons, the wall will allow access to the detector in the garage position.
To improve this study, the PACMAN geometry is recommended to be included in the SiD geometry, which will effect not only the simulations of the muon spoiler background but also all other machine backgrounds.
Finally, the BDS muon background will be used for further occupancy studies together with different other background sources to see the overall occupancy from all background sources.

\section*{Acknowledgments}
Grateful acknowledgments to the SiD Optimization group for their support in performing this study.


\newpage
\bibliographystyle{unsrt}
\bibliography{bibliography.bib}

\end{document}